\documentclass[preprint,osa,showpacs]{revtex4-1}

\usepackage{graphicx}
\usepackage{amsmath,amssymb}
\usepackage{color}
\usepackage{bm}
\usepackage{epstopdf}

\newcommand{\Brho}{\boldsymbol\rho}

\renewcommand{\Im}{{\mathrm {Im}}}
\renewcommand{\Re}{{\mathrm {Re}}}

\newcommand{\curl}{\nabla \times}

\newcommand{\Br}{{\bf r}}

\newcommand{\Bq}{{\bf q}}

\newcommand{\z}{{\bf \hat z}}

\newcommand{\BE}{{\bf E}}

\newcommand{\Tr}{\mathrm{Tr}}

\newcommand{\hvv}[1]{\hat{\boldsymbol{\mathrm{#1}}}}
\newcommand{\vv}[1]{\boldsymbol{\mathrm{#1}}}

\begin{document}

\title{Polarization oscillations of near-field thermal emission}

\author{Manabu Machida}
\email{mmachida@umich.edu}
\affiliation{Department of Mathematics, University of Michigan, Ann Arbor, MI 48109, USA}

\author{Evgenii Narimanov}
\email{evgenii@purdue.edu}
\affiliation{Department of Electrical and Computer Engineering, Purdue University, West Lafayette, IN 47906, USA}

\author{John C. Schotland}
\email{schotland@umich.edu}
\affiliation{Department of Mathematics and Department of Physics, University of Michigan, Ann Arbor, MI 48109, USA}

\date{\today}

\begin{abstract}
We consider the polarization of thermal emission in the near-field of various materials including dielectrics and metallic systems with resonant surface modes. We find that at thermal equilibrium, the degree of polarization exhibits spatial oscillations with a period of approximately half the optical wavelength, independent of material composition. This result contrasts with that of Setala, Kaivola and Friberg [Phys. Rev. Lett. {\bf 88}, 123902 (2002)], who find monotonic decay of the degree of polarization for systems in local thermal equilibrium.
\end{abstract}


\maketitle


The theory of optical coherence is one of the cornerstones of optical physics~\cite{Mandel_Wolf}. Its goal is to describe the statistical regularities of electromagnetic fields in terms of field correlations and their relation to measurable quantities. Implicit in the formulation of coherence theory is the notion of a statistical ensemble of random fields. However, the origin of the randomness of the fields is not explicitly part of the theory.
As a result, the theory of coherence is primarily concerned with the propagation of correlation functions, which has the powerful consequence that its predictions are, in some sense, independent of the underlying probability distribution of the fields.

A notable exception to the above remarks is provided by the theory of thermal emission of radiation~\cite{Mandel_Wolf,James_1991}. In this context, the currents which act as sources of the optical field are taken to obey the fluctuation-dissipation theorem. Thus, it is possible to calculate the correlation function of the field in terms of the statistics of the current. It follows that the emitted field displays both temporal and spatial coherence~\cite{Carter_1975}. Moreover, the near- and far-field coherence functions manifest strikingly different behavior\cite{Carminati_1999,Shchegrov_2000,Joulain_2005,DeWilde_2006,Greffet_2002}. In particular, in materials that support resonant surface waves, the spectrum of emitted radiation changes qualitatively on propagation, ranging from extreme narrowing at subwavelength scales to broad band in the far-field. Likewise, the spatial coherence of emitted light is dramatically modified in the near-field, with a coherence length that is much smaller than the $\lambda/2$ far-field limit of blackbody radiation. In either case, the alteration in coherence is due to the decay of evanescent modes of the field on propagation.

The near-field polarization of thermal emission has also received attention~\cite{Setala_2002_1,Setala_2002_2}. It was found that at local 
\emph{local} thermodynamic equilibrium, the emitted field becomes depolarized, with the degree of polarization decaying monotonically upon propagation into the far-zone. In this Letter, we study the corresponding problem for systems in thermal equilibrium. Instead of monotonic decay, we predict a new physical effect, namely that the degree of polarization exhibits spatial oscillations with a period of approximately $\lambda/2$. We illustrate this effect for several materials including lossless dielectrics and metallic systems with resonant surface plasmon modes.

The fundamental quantity of coherence theory in the space-frequency domain is the cross-spectral density $W_{ij}$, which is defined by
\begin{equation}
W_{ij}(\Br,\Br';\omega)\delta(\omega-\omega')=\langle E_i(\Br,\omega)E_j^*(\Br',\omega')\rangle \ .
\end{equation}
Here $\BE(\Br,\omega)$ is the electric field at the position $\Br$ and frequency 
$\omega$, the presence of the delta function indicates that the field is taken to be statistically stationary and $\langle\cdots\rangle$ denotes the ensemble average. We 
note that $W_{ij}(\Br,\Br';\omega)$ for $\Br\neq \Br'$ is a measure of the spatial 
coherence of the electric field. The degree of coherence is defined to be~\cite{Tervo}
\begin{equation}
\mu^2(\Br,\Br';\omega) = \frac{\Tr\left[W(\Br,\Br';\omega)W(\Br',\Br;\omega)\right]}{\Tr W(\Br,\Br;\omega)\Tr W(\Br',\Br';\omega)} \ .
\end{equation}
It can be seen that $0 \le\mu\le 1$. The case $\mu=0$ corresponds to an incoherent field and $\mu=1$ to a coherent field; otherwise the field is said to be partially coherent.

There is a fundamental link between polarization and coherence. If we consider
$W_{ij}(\Br,\Br';\omega)$ for $\Br=\Br'$, then $W$ is a measure of the polarization of the field. The degree of polarization $P$ is defined as~\cite{Setala_2002_1,Setala_2002_2,Barakat} 
\begin{equation}
P^2(\Br,\omega) =
\frac{3}{2}\left[\frac{\mathop{\mathrm{Tr}}[W^2(\Br,\Br;\omega)]}
{\mathop{\mathrm{Tr}}^2[W(\Br,\Br;\omega)]}-\frac{1}{3}\right] \ .
\label{polarization}
\end{equation}
It can be shown that and $0\le P \le 1$. When $P=0$ the field is said to be unpolarized and if $P=1$ the field is polarized; otherwise the field is partially polarized.

\begin{table}[t]
\caption{}
\begin{center}
\begin{tabular}{lcc}
\hline
Material & $\lambda$ & $\varepsilon$ \\
\hline
Glass & 500 nm \ & \ 2.25 \\
W & 500 nm \ & \ $4.2 + 18.1i$ \\
Ag & 620 nm \ & \ $-15.0 + 1.0i$ \\
SiC & 11.36 $\mu$m \ & \ $-12.2 + 0.07i$ \\
\end{tabular}
\end{center}
\label{table1}
\end{table}

We consider a system consisting of two homogeneous half-spaces which are at thermal equilibrium at a temperature $T$. The lower half-space $z<0$ is taken to consist of
a nonmagnetic lossy material with a generally complex dielectric permittivity $\varepsilon_1(\omega)$. The upper half-space $z>0$ consists of a nonmagnetic material with a real and frequency-independent permittivity $\varepsilon_2$.
In this setting, the fluctuation-dissipation theorem can be used to relate the
cross-spectral density to the Green's tensor $G_{ij}$ by the formula
\begin{equation}
W_{ij}(\Br,\Br';\omega)=2\pi k_0^2\hbar\coth{\left(\frac{\hbar\omega}{2k_BT}\right)}\Im G_{ij}(\Br,\Br') \ ,
\end{equation}
where $k_0=2\pi/\lambda$ is the free-space wavenumber and $k_B$ is Boltzmann's constant~\cite{Agarwal_1975}.

\begin{figure}[t]
\includegraphics[width=0.6\textwidth]{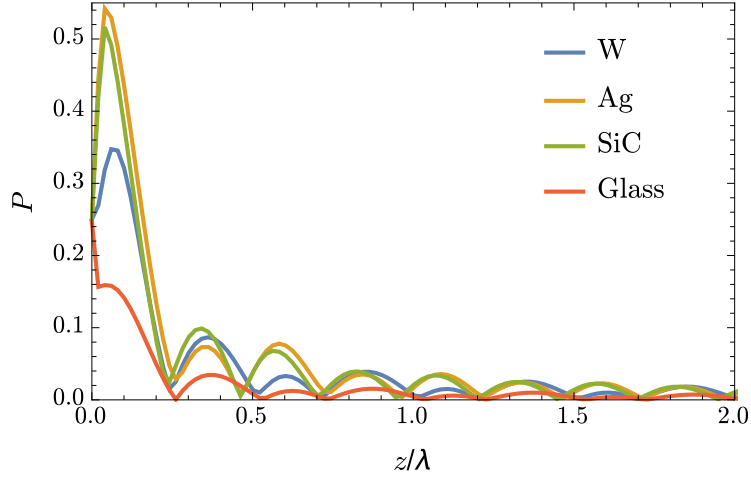}
\caption{(Color online) Degree of polarization of glass, SiC and silver as a function of distance $z$ from the interface. 
\label{fig1}}
\end{figure}

The Green's tensor obeys the equation
\begin{equation}
\curl\curl G(\Br,\Br') -k_0^2\varepsilon(z) G(\Br,\Br')= 4\pi\delta(\Br-\Br')I \ ,
\end{equation}
where
\begin{equation}
\varepsilon(z) =
\begin{cases}
\varepsilon_1 \quad {\rm if}\quad   z < 0 \ , \\
\varepsilon_2 \quad {\rm if}\quad   z > 0  \ ,
\end{cases}
\end{equation} 
and $I$ is the unit tensor. 
The Green's tensor also obeys the boundary conditions
\begin{eqnarray}
\z \times G(\Br,\Br')|_{z=0^+} &=& \z \times G(\Br,\Br')|_{z=0^-} \ , \\
\z \times \curl G(\Br,\Br')|_{z=0^+} &=& \z \times \curl G(\Br,\Br')|_{z=0^-} \ .
\end{eqnarray}
It will prove useful to expand the Green's tensor into plane waves of the form
\begin{equation}
G_{ij}(\Br,\Br') =  \int \frac{d^2q}{(2\pi)^2}e^{i\Bq\cdot(\Brho-\Brho')}
g_{ij}(z,z';\Bq)  \ ,
\label{green}
\end{equation}
where
\begin{eqnarray}
g_{ij}(z,z';\Bq)&=&
\frac{2\pi i}{k_{2z}}\Bigl[
\left(r_s\hat{s}_i\hat{s}_j+r_p\hat{p}_{+i}\hat{p}_{-j}\right)
e^{i k_{2z}(z+z')}
\nonumber \\
&+&
\left(\hat{s}_i\hat{s}_j+\hat{p}_{+i}\hat{p}_{+j}\right)
e^{i k_{2z}(z-z')}
\Bigr] \ .
\label{greenij1}
\end{eqnarray}
Here
\begin{equation}
\hvv{s}=\hvv{q}\times\hvv{z},\quad
\vv{k}_{\pm}=\vv{q}\pm k_{2z}\hvv{z}\ ,\quad
\hvv{p}_{\pm}=\hvv{s}\times\hvv{k}_{\pm}\ ,
\end{equation}
where $k_{2z}=\sqrt{\varepsilon_2k_0^2-q^2}$.  
The Fresnel reflection coefficients are given by
\begin{equation}
r_s=\frac{k_{2z}-k_{1z}}{k_{2z}+k_{1z}} \ ,\quad
r_p=
\frac{k_{2z}\varepsilon_1-k_{1z}}{k_{2z}\varepsilon_1+k_{1z}}\ ,
\end{equation}
where $k_{1z}=\sqrt{\varepsilon_1 k_0^2-q^2}$.

\begin{figure}[t]
\includegraphics[width=0.6\textwidth]{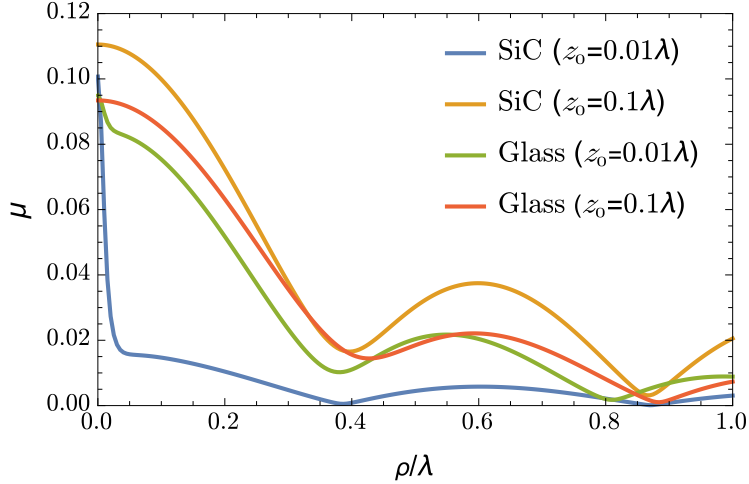}
\caption{(Color online) Degree of coherence of glass and SiC as a function of transverse distance $\rho$ in the plane $z=z_0$ at various distances $z_0$ above the interface. 
\label{fig2}}
\end{figure}

\begin{figure}[t]
\includegraphics[width=0.6\textwidth]{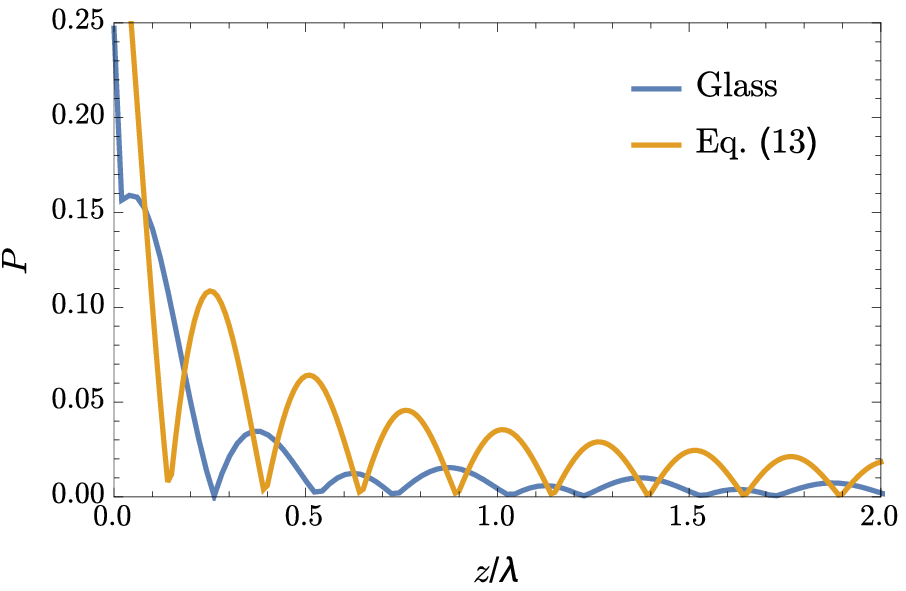}
\includegraphics[width=0.6\textwidth]{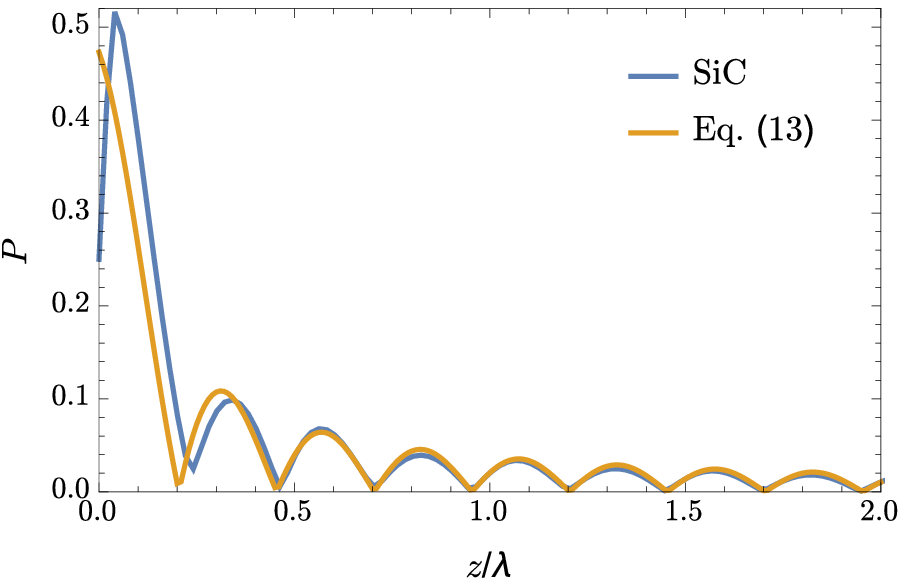}
\caption{(Color online) Comparison of the asymptotic formula (\ref{asymptotics}) 
with the exact result (\ref{polarization}) for the degree of polarization.
\label{fig3}}
\end{figure}

We now describe applications of the above results to various materials. In the cases considered, the system has the temperature $T=300$K and the upper half-space is taken to have permittivity $\varepsilon=1$. The optical properties of the materials that comprise the lower half-space are given in Table~\ref{table1}~\cite{Palik}. In Fig.~\ref{fig1} we plot the degree of polarization $P$ as a function of the distance $z$ from the interface for several materials including glass, tungsten, silver and silicon carbide (SiC). We see that the field is partially polarized near the interface and becomes depolarized at large distances. In Fig.~\ref{fig2}, we plot the corresponding degree of coherence $\mu$ as a function of the transverse distance $\rho=|\Br-\Br'|$, for points $\Br$ and $\Br'$ in the plane $z=z_0$. As may be expected, the field is partially coherent near the interface and becomes incoherent with propagation, consistent with the results of \cite{Carminati_1999}.

Evidently, the behavior of both $P$ and $\mu$ depends sensitively on the dielectric
permittivity of the material under investigation. In the case of glass, which is a non-lossy dielectric at the wavelength $\lambda=500$ nm, the field is relatively less polarized, due to the decay of evanescent modes in the near-field of the interface.
In contrast, silver exhibits a surface plasmon resonance at $\lambda=620$ nm and the near-field is strongly polarized. This should be compared with the case of tungsten, which does not exhibit a plasmon resonance at $\lambda=500$ nm, where it can be seen that $P$ is correspondingly reduced. Finally, we consider the case of SiC, which at $\lambda=11.36$ $\mu$m supports surface-phonon polariton modes. We see that as in the example of silver, the near-field is strongly polarized.

A striking feature of Fig.~\ref{fig1} is the oscillatory nature of the distance-dependence of the degree of polarization. A straightforward asymptotic analysis of the integral (\ref{green}) defining $\Im G_{ij}$ shows that for $z\gg\lambda$, the envelope of the oscillation decays as $P\sim 1/z$. Moreover, if $\Re(\varepsilon) \ll -1$, it can be seen that
\begin{equation}
P\sim \frac{1}{2}\left|\frac{\sin\left(4\pi z/\lambda + 2/\sqrt{\Re(\varepsilon)}\right)}{4\pi z/\lambda + 2/\sqrt{\Re(\varepsilon)}}\right| \ . 
\label{asymptotics}
\end{equation}
Thus, the period of the oscillations is asymptotically  $\lambda/2$, independent of the material. In Fig.~\ref{fig3}, we compare the above asymptotic formula with the exact result obtained from~(\ref{polarization}). As may be expected, there is excellent agreement for SiC and relatively poor agreement for glass.

It is important to note that polarization oscillations are not present for systems in local thermal equilibrium, where $P$ decays monotonically~\cite{Setala_2002_1}. This difference can be explained by the interference between modes in the upper half-space, a mechanism which is not present in the calculations presented in~\cite{Setala_2002_1}. 

In conclusion, we have investigated the polarization of near-field thermal emission. We find that at thermal equilibrium, the degree of polarization exhibits novel spatial oscillations with a period of approximately $\lambda/2$, independent of material composition. The effect is most pronounced in systems with resonant surface waves such as surface-plasmons or surface-phonon polaritons.

We are grateful to Remi Carminati for valuable discussions. This work was supported by the NSF Center for Photonic and Multiscale Nanomaterials.

\end{document}